\documentclass{an}
\usepackage{graphicx}
\usepackage{times}
\begin{document}

\hyphenation{he-re-af-ter Ca-ta-lo-gue the-se me-thod}

\newcommand{\Mo}{$\mathrm{M_\odot}$}
\newcommand{\arcs}{$^{\prime\prime}$}
\newcommand\aaps{{A\&AS}}%
\newcommand\aap{{A\&A}}%
\newcommand\bain{{Bull.~Astron.~Inst.~Netherlands}}%

\Pagespan{1}{}
\Yearpublication{2007}%
\Yearsubmission{2007}%
\Month{}%
\Volume{}%
\Issue{}%

\title{Combining astrometry with the light-time effect: \\
       The case of VW~Cep, $\zeta$~Phe and HT~Vir. \thanks{Based on observations secured at the South Africa
         Astronomical Observatory, Sutherland, South Africa}}

\author{P. Zasche\thanks{\email{zasche@sirrah.troja.mff.cuni.cz}\newline}
\and  M. Wolf}

\titlerunning{Combining astrometry with the light-time effect}
\authorrunning{P. Zasche \& M. Wolf}
\institute{ Astronomical Institute, Faculty of Mathematics and
Physics, Charles University Prague, CZ-180 00 Praha 8, V
Hole\v{s}ovi\v{c}k\'ach 2, Czech Republic}

\received{Today} \accepted{Today} \publonline{later}


\abstract{Three eclipsing binary systems with astrometric orbit
have been studied. For a detailed analysis two circular-orbit
binaries (VW~Cep and HT~Vir) and one binary with an eccentric
orbit ($\zeta$~Phe) have been chosen. Merging together astrometry
and the analysis of the times of minima, one is able to describe
the orbit of such a system completely. The $O-C$ diagrams and the
astrometric orbits of the third bodies were analysed
simultaneously for these three systems by the least-squares
method. The introduced algorithm is useful and powerful, but also
time consuming, due to many parameters which one is trying to
derive. The new orbits for the third bodies in these systems were
found with periods $30$, $221$, and $261$~yr, and eccentricities
$0.63$, $0.37$, and $0.64$ for VW~Cep, $\zeta$~Phe, and HT~Vir,
respectively. Also an independent approach to compute the
distances to these systems was used. The use of this algorithm to
VW~Cep gave the distance $d=(27.90~\pm~0.29)~\mathrm{pc}$, which
is in excellent agreement with the previous \emph{Hipparcos}
result.}

\maketitle

\section{Introduction}

During the last century, many observations of close binary systems
were collected, especially for eclipsing binaries. Most
observations of these stars were made photometrically but in some
cases also the spectroscopy was obtained. In a few cases, also
astrometric observations were carried out and these systems, which
were studied by several different techniques, are the most
interesting, because one is able to find more relevant parameters
of them. Especially, if it is possible to analyse the different
measurements together (it means from different observational
techniques in one least-squares fit) one is able to find the
complete set of parameters describing the orbit of such a binary,
which are not in contradiction. Such approach is potentially very
powerful, especially in upcoming astrometric and photometric space
missions.

There are a few well-known eclipsing binaries with astrometric
measurements for which the light-time effect (hereafter LITE) was
considered, or expected. However, the eclipsing binary (hereafter
EB) nature and the astrometric orbit were usually studied
separately.

For example, there are several published light-curve solutions for
V505~Sgr with the third light included (see e.g. L\'azaro et al.
(2006), {\.I}bano{\v g}lu et al. (2000), etc), and an analysis of
its apparent orbital period changes interpreted as the LITE due to
the third body (e.g. Rovithis-Livaniou et al. 1991). The only
paper which compares the astrometry and a period analysis of $O-C$
deviations from the constant orbital period was published by Mayer
(1997). Despite existing astrometric measurements, there were no
attempts to combine these two methods together. The results from
different approaches were just compared to each other.

Another systems are for instance$\,$ QS~Aql,$\,$ 44i~Boo,$\,$
QZ~Car, SZ~Cam, GT~Mus, or V2388~Oph. The coverage of the
astrometric orbit is very poor for some of them. In the system
QS~Aql the LITE could be determined precisely, but the astrometric
orbit is covered only very poorly, see e.g. Mayer (2004). The
opposite case is V2388~Oph, where the astrometric-orbit parameters
were computed very precisely, but there are only a few
minimum-time measurements (see e.g. Yakut et al. 2004). For SZ~Cam
only a few usable astrometric observations were obtained, but the
LITE is well-defined and also the third light in the light-curve
solution was detected (Lorenz et al. 1998). QZ~Car is a more
complicated, probably quadruple system consisting of an eclipsing
and a non-eclipsing binary, but there are also only a few usable
astrometric measurements. Also GT~Mus is a quadruple system,
consisting of an eclipsing and RS~CVn component. In many other
cases, only measurements of the times of minima are available,
without astrometry. For some others, astrometry without
photometry, is only available. Other systems where the astrometric
observations were obtained and the LITE is observable or expected
are listed in Mayer (2004).

The only paper on combining these two different approaches into
one joint solution is that by Ribas et al. (2002), where a similar
method as described in this paper was applied to the system R~CMa,
but where only a small arc of the astrometric orbit was available.
Besides the astrometry and LITE also the proper motion on the long
orbit was analyzed. On the other hand one has to note, that in
Ribas et al. (2002) the complete astrometric parameters (with
proper motions, parallax, etc.) were used, while in this paper
only a relative astrometry of the distant body relative to the
eclipsing pair was analysed.

\section{Methods}

\subsection{Astrometry}

The number of visual binaries with astrometric orbits has grown,
but the whole orbit is covered with data points only in some
cases. Thanks to precise interferometry the observable semimajor
axes of astrometric binaries are still decreasing down to milli-
and micro-arcseconds.On the other hand, one has to regret that no
recent astrometric measurement of a wide pair of about 1\arcs$\,$
has been obtained for the systems mentioned below. $\,$ All the
astrometric observations were adopted from 'The Washington
Double Star Catalogue' WDS\footnote{http://ad.usno.navy.mil/wds/} (Mason et al. 2001). 

From astrometric data ($N$ measurements of the position angle
$\theta_i$, separation $\rho_i$, the uncertainties
$\sigma_\theta$, and $\sigma_\rho$ and the time of the observation
$t_i$) one is trying to find out the parameters of the orbit for
the distant component, defined by 7 parameters: period $p_3$,
angular semimajor axis $a$, inclination $i$, eccentricity $e$,
longitude of the periastron $\omega_3$, the longitude of the
ascending node $\Omega$, and the time of the periastron passage
$T_0$. One has to solve the inverse problem
$$
    \{ (t_i, \theta_i, \rho_i, \sigma_{\theta,i}, \sigma_{\rho,i}) \}_{i=1,N} \rightarrow ( a, p_3, i, e, \omega_3, \Omega, T_0 ).
    \nonumber
$$
The least-squares method and the simplex algorithm was used (see
e.g. Kallrath \& Linnell 1987).

Comparing the theoretical position on the sky $\theta_0$ and
$\rho_0$ with the observed ones $\theta_i$ and $\rho_i$, one can
calculate the sum of normalized residuals squared
\begin{equation}
  {\chi^2_{astr}} = \sum_{i=1}^N \left[ \left( \frac{\theta_i-\theta_{0,i}}{\sigma_{\theta,i}} \right)^2 + \left( \frac{\rho_i-\rho_{0,i}}{\sigma_{\rho,i}} \right)^2 \right],
  \label{eq7}
\end{equation}
following Torres (2004). With this $\chi^2_{astr}$ and using the
simplex algorithm one can obtain a set of parameters ($a$, $p_3$,
$i$, $e$, $\omega_3$, $\Omega$, $T_0$) describing the astrometric
orbit.

The weighting is provided by the uncertainties $\sigma_\theta$ and
$\sigma_\rho$. These values are obtained from the observations, or
estimated as some typical uncertainty level for the certain kind
of measurement provided by a specific instrument.

The efficiency and the computing time required by the algorithm
strongly depends on the initial set of parameters and the chosen
trial step in the parameter space. If nothing is known about the
solution, one has to scan a wide range of parameters (eccentricity
$e$ from $0$ to $1$, and the angular parameters from $0^\circ$ to
$360^\circ$).

The efficiency of the algorithm could be improved if the simplex
is used repeatedly. It can happen that the simplex converges into
a local minimum while the global one is far away. It is therefore
advisable to run the algorithm again, with as large initial steps
as in the previous run, but keeping the values of the parameters
corresponding to the previously found minimum as one vertex.
Repeating this strategy several times over the whole parameter
space, one can judge whether the global minimum was found by
checking whether the sum of squares of the residuals is still
changing or not.

To conclude, using this strategy and the combined method described
in section \ref{comb}, one gets the satisfactory result after
large number of iterations. This number and computing time is
strongly depended on the input data set and the number of
parameters fitted. For the case VW~Cep with the largest data set
(about 1600 data points, see below) and also with the most
parameters to fit (14 in total) one reaches the solution, when the
sum of squares is not changing significantly, after circa 100,000
simplex steps. This takes about one day on computer with 2 GHz
processor.

\subsection{Light-time effect}

A different method to study eclipsing binaries in hierarchical
triple systems is based on the eclipse timings. The light-time
effect (or the 'light-travel time') causes apparent changes of the
observed binary period with a period corresponding to the orbital
period of the third body. This useful method is known for decades
and its detailed description was presented by Irwin in 1959, while
some comments on it and its limitations were discussed by
Frieboes-Conde \& Herczeg (1973) and by Mayer (1990).

From the numerical point of view the method is quite similar to
the previous one, because it is also an inverse problem. One has
$M$ measurements of the times of minima of the system at certain
constant $JD_i$ with the individual uncertainties $\sigma_m$. The
task is to find five parameters describing the orbit of the third
body in the system: the period of the third body $p_3$, the
semiamplitude of the LITE $A$, the eccentricity $e$, the time of
the periastron passage $T_0$, and the longitude of periastron
$\omega_{12}$. One has to compute simultaneously also two (or
three) parameters of the eclipsing binary itself, namely its
linear (or quadratic) ephemeris $JD_0$ and period $P$ (and 
$q$ for the quadratic one). Altogether, one has 7 (or 8)
parameters to derive from the model fit of the minimum-time
measurements
$$
 \{ (JD_i, \sigma_{m,i}) \}_{i=1,M} \rightarrow (p_3, A, e, T_0, \omega_{12}, JD_0, P, q).
 \nonumber
$$
Similarly as in the astrometry case, the resultant sum of
normalized square residuals is
  \begin{equation}
    \chi^2_{LITE} = \sum_{i=1}^M \left( \frac{(O-C)_i}{\sigma_{m,i}} \right) ^2,
    \label{eq11}
  \end{equation}
where $O$ and $C$ stand for observed and computed time of minimum,
respectively. The same remarks about the efficiency and the
computing time required discussed for the astrometric solution
also apply here.

At this place it is necessary to remark one useful comment. One
has to distinguish between the two angles $\omega_3$ and
$\omega_{12}$. The parameter used in LITE analysis is
$\omega_{12}$, but in astrometry the quantity
$\omega_3~=~\omega_{12}~+~\pi$ is employed. In this paper the
angle $\omega$ stands for the longitude of the periastron for the
eclipsing binary, i.e. $\omega~=~\omega_{12},$ and the subscripts
will be omitted for clarity.

\subsection{Combining the methods} \label{comb}

The task is to combine the astrometry and the analysis of times of
minima into one joint solution together. Having $N$ astrometric
and $M$ minimum time measurements, one is able to merge them
together and obtain a common set of parameters
  \begin{eqnarray}
 (t_i,\theta_i,\rho_i,&\!\!\sigma_{\theta,i}&\!\!,\sigma_{\rho,i},JD_i,\sigma_{m,i}) \rightarrow  \nonumber \\
 & \rightarrow & (a,p_3,i,e,\omega,\Omega,T_0,JD_0,P,q).
  \nonumber
  \end{eqnarray}
This set of 10 parameters fully describes the orbit of the
eclipsing binary around the common centre of mass with the third
unresolved component together with the ephemeris of the binary
itself. It is necessary to solve one least-squares fit of these 10
parameters.

One is able to find out the mass of the third body and the
semimajor axis of the wide orbit because the inclination is known
and one can calculate the mass function of the wide orbit
  \begin{eqnarray}
    f(M_3) & = & \frac {(a_{12}\sin i)^3} {p_3^2} = \frac {(M_3 \sin i)^3} {(M_1+M_2+M_3)^2} =   \nonumber \\
           & = &  \frac {1}{p_3^2} \cdot {\left[ \frac
           {173.15 \cdot A} {\sqrt{1-e^2\cos^2\omega}}
           \right]^3},
    \label{eq13}
  \end{eqnarray}
where $a_{12}$ stands for the semimajor axis of the binary orbit
around the common centre of mass and $M_1, M_2, M_3$ are the
masses of the primary, secondary, and tertiary component,
respectively. For more details see e.g. Mayer (1990).

The only difficulty which remains unclear is the connection
between the angular semimajor axis $a$ and the amplitude of LITE
$A$. The quantity $a_{12}$ could be derived from Eq.~\ref{eq13}
and with the masses of the individual components one is able to
calculate also the value $a_3$, i.e. the semimajor axis of the
third component around the barycentre of the system
  \begin{equation}
     a_3= a_{12} \cdot \frac{M_1 + M_2}{M_3}.
  \end{equation}
The total mutual distance of the components is $a_\mathrm{total} =
a_{12} + a_3$. Using the \emph{Hipparcos} parallax $\pi$ (Perryman
\& ESA 1997) one can obtain the distance $d$ to the system. Now it
is possible to enumerate the angular semimajor axis $a$ as a
function of $d$ and $a_\mathrm{total}$
  \begin{equation}
    a~=~\arcsin \left( \frac{a_\mathrm{total}}{d} \right).
  \end{equation}
The way how these two different approaches were putted together is
following the similar approach by Torres (2004). From the
mathematical point of view both methods are analogous and there is
an overlap of the parameters in both methods. Our task is to
minimize the combined $\chi^2$
  \begin{equation}
   \chi^2_{comb} = \chi^2_{astr} +  \chi^2_{LITE},
  \end{equation}
where the values $\chi^2_{astr}$ and $\chi^2_{LITE}$ are taken
from the Eqs.\ref{eq7} and \ref{eq11}.

Sometimes there arises a problem with the $\chi^2$ values in both
methods, which are incomparable. Especially when there are much
more data points in one method against the other, also the
resultant $\chi^2$ would be much larger and as a consequence this
method outweigh the other one. This problem could be eliminated
using new uncertainties $\sigma'$ 
$$ \sigma_{\theta}' = \sqrt{N} \cdot \sigma_{\theta},
\,\,\,\, \sigma_{\rho}' = \sqrt{N} \cdot \sigma_{\rho}, \,\,\,\,
\sigma_{m}' = \sqrt{M} \cdot \sigma_{m} \, .$$

\section{An application of the method to several particular systems}

The method described above was tested on a few systems satisfying
the following conditions: 1. More than 10 times of minima and more
than 10 astrometric observations are available. 2. The observed
range of the position angle in the astrometric measurements is
larger than 10$^\circ$.

\begin{table*}
\caption{The final results: Parameters of the third bodies from
combined astrometry and the light-time effect. The table is
divided into three parts, in the first one are ten computed
parameters, in the second one the values from the literature and
in the last one the quantities computed from the previous parts.
The values of parallax and distance were adopted from the
\emph{Hipparcos} measurements. In the row 'Data set', 'a' and 'm'
denote the number of astrometric observations, and times of
minima, respectively.}
 \centering
\begin{tabular}{c c c c c }
\hline\hline
Parameter   &  Unit    &            VW~Cep            &         $\zeta$~Phe       &        HT~Vir            \\
\hline
  $JD_0$   & $[$HJD$]$ & $2437001.4327 \pm 0.0025$    & $2441643.7382 \pm 0.0008$ &  $2452722.5040 \pm 0.0050$ \\
  $P$      & $[$d$]$   & $0.278315349 \pm 0.00000012$ & $1.6697772 \pm 0.0000013$ &  $0.4076696 \pm 0.0000025$ \\
  $p_3$    & $[$yr$]$  & $29.79 \pm 0.08$             & $220.9 \pm 3.5$           &  $260.7 \pm 0.40$         \\
  $T_0$    & $[$HJD$]$ & $2450390.6 \pm 37.3$         & $2419908.9 \pm 2465.6$    &  $2442832.3 \pm 60.6$    \\
  $\omega$ & $[^\circ]$& $239.28 \pm 2.88$            & $97.1 \pm 2.2$            &  $250.9 \pm 0.6$        \\
  $e$      &           & $0.633 \pm 0.007$            & $0.366 \pm 0.082$         &  $0.640 \pm 0.005$     \\
  $A$      & $[$d$]$   & $0.0131 \pm 0.0004$          & $0.0808 \pm 0.0080$       &  $0.1274 \pm 0.0024$  \\
  $a$      & $[$mas$]$ & $445.2 \pm 33.0$             & $859.5 \pm 137.2$         &  $1023.5 \pm 183.3$   \\
  $i$      & $[^\circ]$& $33.6 \pm 1.2$               & $64.4 \pm 3.0$            &  $45.4 \pm 3.7$      \\
  $\Omega$ & $[^\circ]$& $17.7 \pm 3.1$               & $33.5 \pm 4.9$            &  $180.8 \pm 2.6$     \\
  \hline
  $M_{12}$ & [\Mo]     & $1.37$                       & $6.48$                    &  $2.3$               \\
 References &          & Kaszas et al. 1998           & Andersen 1983             &  D'Angelo et al. 2006 \\
  $\pi$    & $[$mas$]$ & $36.16 \pm 0.97$             & $11.66 \pm 0.77$          &  $15.39 \pm 2.72$     \\
  $d$      & $[$pc$]$  & $27.7 \pm 0.7$               & $85.8 \pm 5.7$            &  $65.0 \pm 11.5$     \\
  \hline
  $a_{12}$ & $[$AU$]$  & $4.33 \pm 0.20 $             & $15.52 \pm 1.60$          &  $32.23 \pm 0.74$     \\
  $f(M_3)$ & [\Mo]     & $0.0154 \pm 0.0016$          & $0.0562 \pm 0.0167$       &  $0.1692 \pm 0.0098$ \\
  $M_3$    & [\Mo]     & $0.74 \pm 0.07 $             & $1.73 \pm 0.26$           &  $2.10 \pm 0.10$    \\
  \hline
  Data set &           & 16a + 1594m                  &  12a + 36m                &  275a + 31m    \\
\hline
\end{tabular}
  \label{Table1}
\end{table*}

\subsection{VW~Cep}

The eclipsing binary VW~Cep (HD 197433, BD +75 752, HIP 101750, $V
= 7.3$ mag, sp G5V + G8V) is a W~UMa system and in fact it is one
of the most often observed and analysed system. Both components
are chromosphericaly active. VW~Cep is rather atypical, because
during the primary (the deeper one) eclipse is the less massive
star (the hotter one) occulted by the larger companion (the more
massive and the cooler one).

The first observations of its light variations were done by Schilt
(1926). Since 1946, a large amount of photoelectric observations
was obtained. However, the observed minima times did not fit the
ephemeris due to the LITE and the mass transfer between
components. There were many light-time effect studies of this
system and Herczeg \& Schmidt (1960) proposed the presence of a
third body with an orbital period of 29 years and an angular
distance of the third component between 0.5~\arcs$\,$ and 1.2
\arcs.

In 1974, the first successful visual observation of the third
component was obtained and since then, there were 16 observations
of it. Regrettably, the observations near the periastron passage
are missing (the gap in data is from 1991 to 1999). The last two
measurements ($\theta = 231.4 ^\circ$, $\rho = 0.702$\arcs~and
$\theta = 232.6 ^\circ$, $\rho = 0.695$\arcs) were not published
yet. These were obtained by a speckle camera in April 2007 and
were kindly sent by Elliot Horch (priv.comm.).

\begin{figure}[b!]
 \centering
 \scalebox{0.43}{\includegraphics{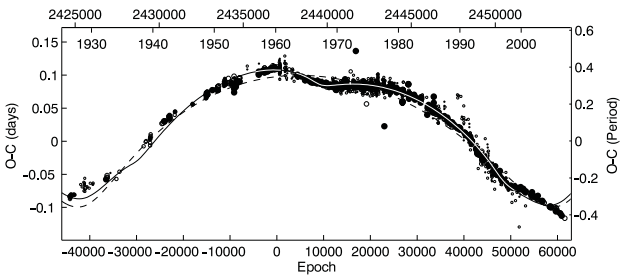}} 
 \caption{An $O-C$ diagram of VW~Cep. The individual observations
 are shown as dots (primary) and open circles (secondary), the
 small ones for visual and the large ones for CCD and photoelectric
 observations. The curves represent the predicted LITE$_3$ +
 LITE$_4$ (the solid one) and only LITE$_4$ (the dashed one).}
 \label{FigVWCep-OC}
\end{figure}

The most complete set of times of minima is in the most recent
period study of VW~Cep by Pribulla (2000). The first times of
minima are from the 1920's and altogether 1907 minima were
collected. From this set of times of minima 313 measurements were
neglected due to their large scatter (mostly the visual ones).
This new minimum-time analysis is based on a larger data set
(about 750 times of minima more than were used by Pribulla et
al.). Two new CCD observations of the minimum light of VW~Cep were
obtained in Ond\v{r}ejov observatory with the 65-cm telescope and
Apogee AP-7 CCD camera, 2 seconds exposure time in R
filter. These new times of heliocentric 
minima are the following: $2454154.3094~\pm~0.0001$ and
$2454195.49866~\pm$ $\pm~0.00012$.

The short-term variations with the period of about two years (see
e.g. Kwee 1966 and Hendry \& Mochnacki 2000) are probably caused
by the surface activity cycles on the primary component. Due to
this activity an unique interpretation of the behaviour of period
changes is still missing. Pribulla et al. proposed a mass transfer
(the quadratic term) plus the third and the fourth body in the
system (two periodic terms). Nevertheless, they were not able to
explain the $O-C$ diagram in detail.

\begin{figure}[b!]
  \centering
  \scalebox{0.43}{\includegraphics{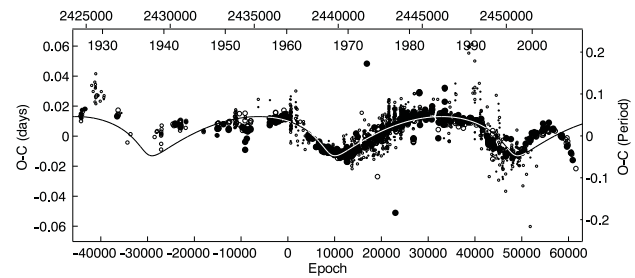}}
  \caption{An $O-C$ diagram of VW~Cep after subtraction of the LITE$_4$.
  The description is the same as in the previous $O~\!-~\!C$ figure, and the
  solid line represents the LITE caused by the third component in the system.}
  \label{FigVWCep-OC2}
\end{figure}

Another approach has been chosen in this paper. Especially due to
only a few astrometric observations (16 measurements from 1974 to
2007) we have decided to explain only the most significant effects
in the $O-C$ diagram. It means the astrometric variation with a
period of about 30 years has been identified with the $O-C$
variation with the same period, but the long-term variation in the
$O-C$ diagram (which has been interpreted by Pribulla et al. as a
quadratic term) was explained as a variation due to the fourth
body on its very long orbit. This approach was chosen especially
because of the systemic-velocity variations, see below. It means
during the computation process the value $\chi^2_{comb}$ was
minimized with respect to 14 parameters in total:\\ $
(A,p_3,i,e,\omega,\Omega,T_0,JD_0,P,A_4,p_4,e_4,\omega_4,T_{0,4}).$

\begin{figure}
  \centering
  \includegraphics[width=8.2cm]{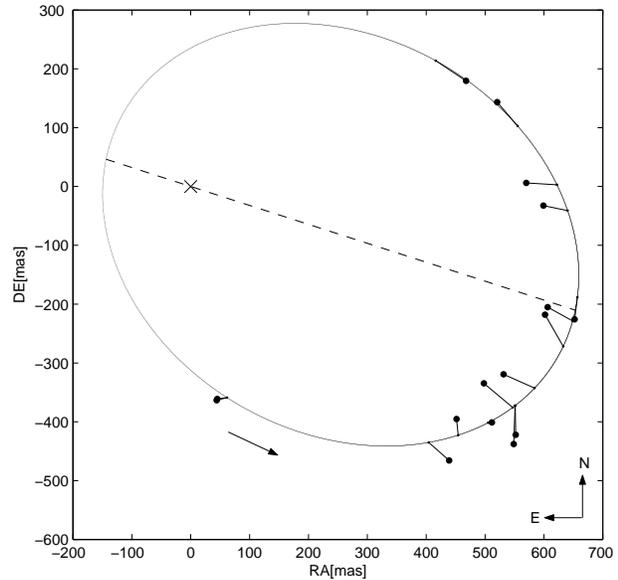}
  \caption{Relative orbit of VW~Cep on a plane of the sky. The points represent
  individual observations, while the solid curve corresponds to the
  solution described in the text. The straight lines connect
  individual observations with their expected positions on the
  fitted orbit. The cross indicates the position of the
  eclipsing binary on the sky and the arrow the direction of
  the orbit of the third body. The dashed-line represents the
  line of apsides.}
  \label{FigVWCep-orbit}
\end{figure}

The combined approach of analysing the times of minima together
with astrometry led to the parameters shown in Table \ref{Table1}
and \ref{Table2}. The $O-C$ diagram in Fig. \ref{FigVWCep-OC}
shows the times of minima together with the curve which represents
the LITE$_3$ + LITE$_4$. If one subtracts only the LITE$_4$
variation and try to describe the behavior of the minimum times,
one gets Fig. \ref{FigVWCep-OC2}, where only the LITE$_3$ caused
by the third component is displayed. The fit is not very
satisfactory because of the presence of the chromospheric activity
of the individual components, or due to a putative additional
component (see e.g. Pribulla 2000). In Fig. \ref{FigVWCep-orbit},
the astrometric orbit of the binary with the individual
measurements and their theoretical positions is shown.
Regrettably, no observations near the periastron passage are
available. The curve also represents the theoretical orbit
according to the parameters given in Table \ref{Table1} in
agreement with the LITE analysis.

The parameters describing the LITE$_3$ and LITE$_4$ variations are
in Table \ref{Table1} and \ref{Table2} and could be compared to
the parameters derived during the previous analysis by Pribulla et
al. Their values for the third-body orbit are: $p_3 = 31.4$~yr,
$e~=~0.77$, $\omega~=~183^\circ$, and
$a_\mathrm{total}~=~12.53$~AU. Our values are in Table
\ref{Table1} except for $a_\mathrm{total}~=~12.35$~AU, and as we
can see they differ significantly in several parameters. This is
due to completely different approach describing the $O-C$
variations. Only the period and the amplitude of such variation
are comparable, but these are the most important for our combined
solution. One has also to disagree with the result by Pribulla et
al., that the astrometric orbit could not be identified with the
LITE$_3$ variation from the $O-C$ diagram. As one can see, our new
results are in agreement with each other without any problems.

\begin{table}
\caption{VW~Cep: parameters of the fourth-body orbit}
 \centering
\begin{tabular}{c c c}
\hline\hline
  Parameter   & Unit & Value $\pm$ Error \\
\hline
   $p_4$      & [yr] & 77.46 $\pm$ 0.04  \\
   $A_4$      & [d]& 0.100 $\pm$ 0.002 \\
   $\omega_4$ & [$^\circ$]& 283.42 $\pm$ 2.30 \\
   $T_{0,4}$  & [HJD]& 2453857.4 $\pm$ 13.6 \\
   $e_4$      &      & 0.543 $\pm$ 0.007  \\
\hline
\end{tabular}
\label{Table2}
\end{table}

Also the astrometric orbit could be compared with the previously
published one. Most recently Docobo \& Ling (2005) 
published the following parameters of the astrometric orbit: $p_3~=~31.0$~yr, $a~=~485$~mas, $i~=~39.3^\circ$, and $e~=~0.68$. 
If one compares these values with the new ones (see Table
\ref{Table1}), one can see that the differences are slightly
beyond the limits of errors.

If we assume the mass of the eclipsing binary to be $M_{12} =
1.37~$\Mo~(Kaszas et al. 1998) and the parallax
$\pi~=~36.16~\mathrm{mas}$ (from Perryman \& ESA 1997), the
distance to the system is only about $27.66~\mathrm{pc}$, which
results in the third-body mass of $M_3~=~0.741~$\Mo. The distant
component is about 2.2 magnitudes fainter than the VW~Cep itself,
so its luminosity and also mass should be much smaller than the
mass of the eclipsing components. The total bolometric magnitude
of VW~Cep is about 4.7 mag, so the magnitude of the third
component is about 6.9 mag, which leads to the spectral type of
about K3. The typical mass of this spectral type is about
$0.75~$\Mo~(according to Harmanec 1988), which is in good
agreement with the new result and within its error limits.

\begin{figure}[b!]
  \centering
  \scalebox{0.95}{
  \includegraphics[width=8.5cm]{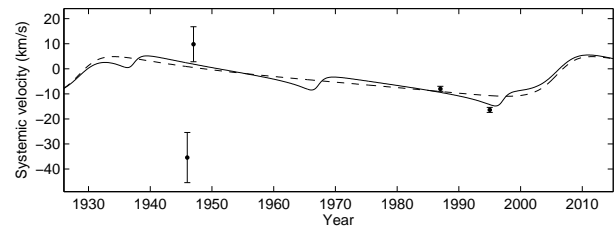}}
  \caption{Systemic velocity variations in VW~Cep. The
  individual points represent computed systemic velocities (see
  details in text). The solid curve stands for the combined
  LITE$_3$ + LITE$_4$ variations, while the dashed one only
  for the LITE$_4$ variation.}
  \label{FigVWCep-RV}
\end{figure}

Different systemic velocities $v_\gamma$ were found at different
epochs. These values are: $v_\gamma = (-35.4~\pm 10)~ \mathrm{km
\cdot s^{-1}}$ (Popper 1948), $(+9.8~\pm 7)~\mathrm{km \cdot
s^{-1}}$ (Binnendijk 1966), $(-8~\pm 1)~\mathrm{km \cdot s^{-1}}$
(Hill 1989), and $(-16.4~\pm 1)~\mathrm{km \cdot s^{-1}}$ (Kaszas
1998). In the time plot (see Fig. \ref{FigVWCep-RV}) one can see
the curve which represents the theoretical variation of $v_\gamma$
caused by the orbital motion around the common barycentre. Except
for the first one data point by Popper the systemic velocities are
following the long-term variation and are almost within its errors
near the theoretical values. The value of Popper is affected by a
relatively large error. The scatter of the individual RV data
points is much larger than that from Binnendijk, which could be
caused by the combination of two different data sets from
different instruments and obtained after more than 600 orbital
revolutions (which could shift the ephemeris). Pribulla \&
Rucinski (2006) suggested that the scatter of the systemic
velocity data points is instrumental, which seems unlikely for
such a large amplitude. For the final confirmation of $v_\gamma$
variations a more accurate and larger data set is necessary.

We also tried to derive the parallax of VW~Cep using this combined
approach. Leaving the parallax as another free parameter, it was
calculated from the comparison of the angular and absolute
semimajor axis. Using this approach, almost all of the relevant
parameters remained nearly the same, only the inclination changed
a bit, being about $1^\circ$ lower and the angle $\Omega$ about
$1.5^\circ$ lower. The new parameters led to a higher third mass
of $M_3 = 0.76~$\Mo. The parallax decreased from
$(36.16\pm0.97)~\mathrm{mas}$ (\emph{Hipparcos}) to
$(35.85~\pm~0.37)~\mathrm{mas}$. This parallax would shift the
distance from $(27.7\pm0.7)$ pc (\emph{Hipparcos}) to
$(27.90~\pm~0.29)$ pc. As one can see, the result of the parallax
is more precise than any of the previously derived parallaxes (see
Hendry \& Mochnacki (2000) for a summary of the previous values).

One could conclude that the third body is probably of spectral
type K3 with a mass around $0.74$~\Mo. It is clear, however, that
a more complicated model will be needed to describe the observed
changes completely.

\subsection{$\zeta$~Phe}

The system $\zeta$~Phe is the brightest eclipsing binary with two
components of early spectral types, exhibiting total and annular
eclipses. This is the only binary with an eccentric orbit included
in this study. $\zeta$~Phe (HD 6882, HR 338, HIP 5348, $V = 4.0$
mag, sp B6V + B9V) is an Algol-type eclipsing binary. It is a
visual triple and SB2 spectroscopic binary.

The brightest component is the EB, while the most distant
component is the faintest one (some 6\arcs away and with its
apparent brightness of about $8$~mag). The last known component is
a $7^\mathrm{th}$-magnitude star at a distance of about
$600~\mathrm{mas}$. This is the astrometric component and in our
opinion also the one which causes the LITE. The depth of the
primary minimum of the eclipsing pair is about $0.5$~mag and the
period is about $1.66$~d.

\begin{figure}
   \centering
   \scalebox{0.426}{\includegraphics{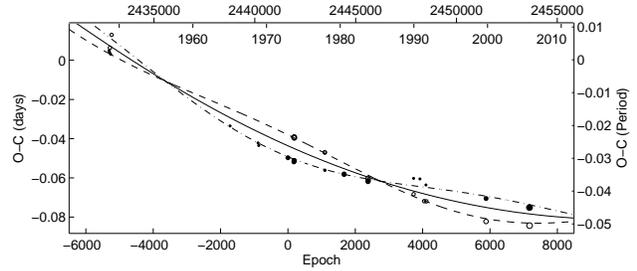}}
   \caption{The $O-C$ diagram of $\zeta$~Phe. The description is the
   same as in the previous $O-C$ figures. The apsidal motion curve
   (the dash-dotted one for the primary and the dashed for the secondary) is
   plotted around the (solid) LITE curve.}
   \label{ZetaPheOC}
\end{figure}

The unfiltered light curve was observed in 1950's by Hogg (1951),
after then by Dachs (1971) in $UBV$ filters, and the best one by
Clausen et al. (1976) in $ubvy$ filters. In this latter paper all
relevant parameters of the eclipsing system were derived and also
the third light was computed. Its value changes from $3\% (u)$ to
$8\% (y)$ and the distant component was classified as a spectral
type A7 star.

Clausen et al. (1976) also collected the times of minima obtained
before 1975. They concluded that no significant apsidal motion is
observed. The first apsidal-motion study was published by
Gim\'enez et al. (1986). With an updated list of the times of
minima one is able to conclude that the apsidal motion is
definitely present. It is clearly visible in the $O-C$ diagram
shown in Fig.~\ref{ZetaPheOC}. Altogether 36 times of minima used
here came from the paper cited above and from Mallama (1981),
Gim\'enez et al. (1986), Kv\'iz et al. (1999). The most recent
ones are in Table \ref{ZetaPheMin}.

Our new photoelectric $UBV$ observations were secured with the
modular photometer utilizing Hamamatsu EA1516 photomultiplier on
the 0.5-m telescope at the Sutherland site of the South African
Astronomical Observatory (SAAO) during two weeks in September
2005. The Johnson $UBV$ photoelectric measurements were secured
with 10-second integration times. Each observation of $\zeta$~Phe
was accompanied by an observation of the local comparison star
$\eta$~Phe ($V$ = 4.36 mag). All measurements were carefully
reduced to the Cousins E-region standard system (Menzies et al.
1989) and corrected for differential extinction using the
reduction program HEC~22 rel. 14 (Harmanec \& Horn 1998). The
standard errors of these measurements were about 0.008, 0.006, and
0.005 magnitude in $U, B$, and $V$ filters, respectively.

\begin{figure}
   \centering
   \includegraphics[width=8.2cm]{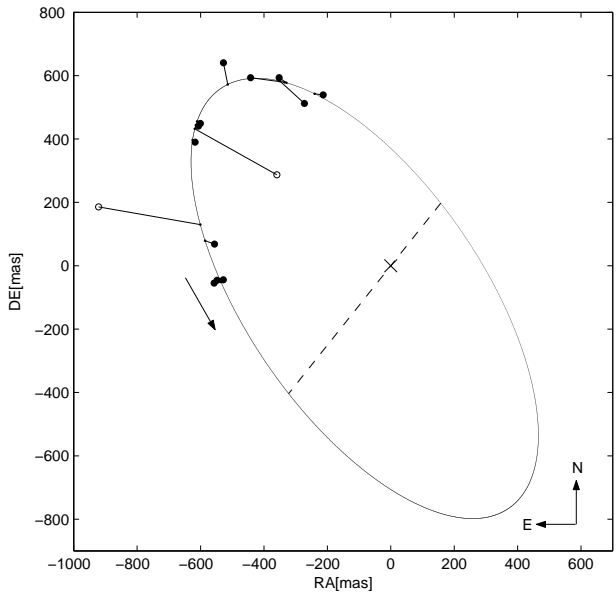}
   \caption{Relative orbit of $\zeta$~Phe on a plane of the sky,
   for a detailed description see Fig. \ref{FigVWCep-orbit}.
   Two measurements (the open circles) were neglected.}
   \label{ZetaPhe-orbit}
\end{figure}

The new times of primary and secondary minimum and their errors
were derived using a least-squares fit to the data and by the
bisecting-chord method. Only the bottom parts of the eclipses were
used. The mean values and the errors for each individual filter
are given. Six new times of minimum light were derived using the
\emph{Hipparcos} photometry (Perryman \& ESA 1997) and fitting the
published light curve. These new times of minima are also included
in Table~\ref{ZetaPheMin}. In this Table, the epochs are
calculated from the light elements given in Table \ref{Table1},
the other columns being self-explanatory.

$\zeta$~Phe has one of the shortest apsidal motions among the
eclipsing binaries (see e.g. Claret \& Gim\'enez 1993). Due to a
low eccentricity, the amplitude of the effect is small. For an
accurate calculation of the apsidal motion rate the method
described by Gim\'enez \& Garcia-Pelayo (1983) was routinely used.
The eccentricity of the orbit in the eclipsing binary is
$e'=0.0107 \pm 0.0020$, the longitude of periastron $\omega_0 =
12.96^\circ \pm 5.96^\circ$, and the apsidal motion rate
$\dot\omega = (0.028 \pm 0.001) \mathrm{^\circ / cycle} = (6.16
\pm 0.20) \mathrm{^ \circ / yr} $, i.e. the apsidal motion period
$U = 58.5$~yr. The most recent apsidal-motion analysis is more
than 20 years old, made by Gim\'enez et al. (1986), but with no
LITE and with a smaller set of times of minima. The eccentricity
derived by Gim\'enez et al. was almost the same, but the apsidal
motion rate $\dot\omega$ was $0.0373~\mathrm{^\circ / cycle}$ and
the angle $\omega_0~=~13^\circ.$

Our approach was a combination of the two different effects. The
behaviour in the $O-C$ diagram was described as a sum of apsidal
motion and LITE contribution $(O-C) = (O-C)_{apsid} +
(O-C)_{LITE}$, distinguishing the primary and secondary minima. It
means the least-squares algorithm was minimizing the
$\chi^2_{comb}$ with respect to 12 parameters in total
($A,p_3,i,e,\omega,\Omega,T_0,JD_0,P,\dot\omega,\omega_0,e'$).

\begin{table}[b!]
\caption{The new minima timings of $\zeta$~Phe based on
photoelectric observations.} \label{ZetaPheMin} \centering
\begin{tabular}{l l c c r c}
\hline\hline
 HJD-2400000 & Error & Prim/Sec & Epoch & Ref.  \\
\hline
 47872.7742 & 0.005  & Sec   & 3730.5  & [1] \\
 47873.6172 & 0.005  & Prim  & 3731.0  & [1] \\
 48187.535  & 0.005  & Prim  & 3919.0  & [1] \\
 48397.0807 & 0.005  & Sec   & 4044.5  & [1] \\
 48484.7523 & 0.005  & Prim  & 4097.0  & [1] \\
 48508.9557 & 0.005  & Sec   & 4111.5  & [1] \\
 51466.9675 & 0.0005 & Prim  & 5883.0  & [2] \\
 51467.7907 & 0.001  & Sec   & 5883.5  & [2] \\
 53622.6453 & 0.0001 & Prim  & 7174.0  & [3] \\
 53623.4710 & 0.0001 & Sec   & 7174.5  & [3] \\
\hline
\end{tabular}
\begin{list}{}{}
\item[Ref.:] [1] - Perryman \& ESA 1997; [2] - Shobbrook (2004);
[3] - This paper;
\end{list}
\end{table}

The astrometric solution based on the combined app-roach is
satisfactory, while the older measurements have larger scatter
than the recent ones (the old ones are visual, while the modern
ones are speckle-interferometric). Two measurements were
neglected, because of their large scatter (see
Fig.\ref{ZetaPhe-orbit}). Our solution led to the parameters
listed in Table~\ref{Table1}, which could be compared to
previously found values. Most recently Ling (2004) reported the
parameters: $p_3~=~210.4$ yr, $e~=~0.348$, $a~=~804$~mas,
$i~=~61.9^\circ$, $\Omega~=~33.5^\circ$, $\omega_3~=~271.7^\circ$.
It is evident that the new parameters are in very good agreement
with these ones. The new values imply the mass function of the
distant body $f(M_3) = 0.056~\mathrm{M_\odot}$. With the masses of
the primary and secondary component of the eclipsing binary $M_1 =
3.93~$\Mo~and $M_2 = 2.55~\mathrm{M_\odot}$ (Andersen 1983), one
could derive the mass of the astrometric third body $M_3 =
1.73~$\Mo. This value corresponds to a spectral type around A7,
which is in excellent
agreement with the photometric analysis (Clausen et al. 1986). 

\subsection{HT~Vir}

One member of the visual binary STF~1781 is the eclipsing binary
system HT~Vir (ADS~9019, HD~119931, HIP~67186, BD+05~2794,
$V~=~7.2$~mag, sp~F8V). HT~Vir is a contact W~UMa system, with a
period of about $0.4$ d and the depths of minima of about
$0.4$~mag. Both visual components have almost equal brightness.
The third component of the system is brighter than the eclipsing
binary HT~Vir during its eclipses and fainter than it during its
maxima.

According to Walker \& Chambliss (1985) the distant astrometric
component was discovered by Wilhelm Struve in 1830 at a separation
of about 1.4~\arcs$\,$ and position angle $240^\circ$. Since then,
numerous astrometric observations were obtained (altogether 277,
from which 275 were used in our analysis) and the orbit is almost
completely covered by the observations.

\begin{figure}
  \centering
  \includegraphics[width=8.2cm]{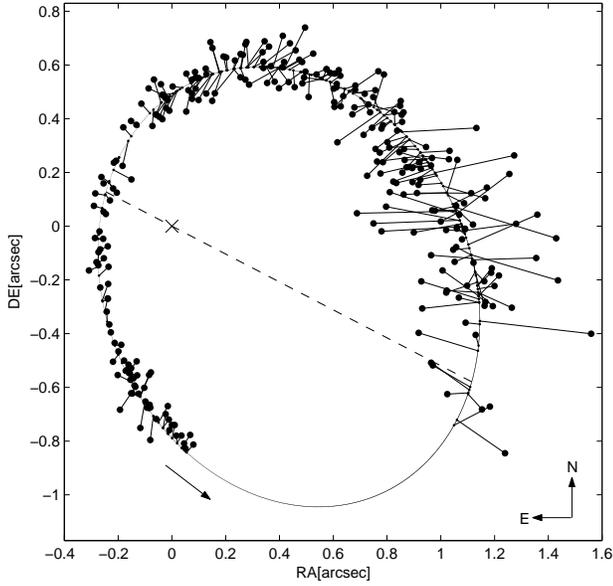}
  \caption{Relative orbit of HT~Vir on a plane of the sky,
  for a detailed description see Fig. \ref{FigVWCep-orbit}.}
  \label{HTVir-orbit}
\end{figure}

Baize (1972) suggested that the star might be variable. After
then, Walker \& Chambliss (1985) obtained a complete light curve
of HT~Vir and did the first analysis. It indicated that both
components of the eclipsing pair are almost identical and in
contact. The temperatures of both components are about
$6000~\mathrm{K}$ and the spectral type is estimated as F8V for
the primary and a little bit earlier for the secondary, the
inclination is close to $90^\circ$. The total mass of the
eclipsing pair is $M_{12}~=~2.3~$\Mo~(D'Angelo et al. 2006).

Lu et al. (2001) discovered that the distant component is also a
binary. They have measured the spectra of the HT~Vir eclipsing
pair, and discovered also the lines from the third component in
the spectra and their RV variations with a period of about $32.45$
d. We therefore deal with a quadruple system.

\begin{figure}
  \centering
  \scalebox{0.426}{\includegraphics{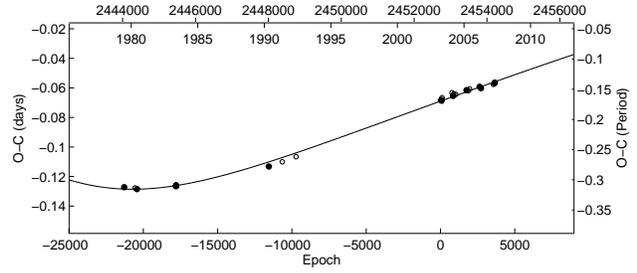}}
  \caption{An $O-C$ diagram of HT~Vir. The description is the
  same as in the previous $O-C$ figures, all minimum times are
  photoelectric or CCD ones.}
  \label{HTVirOC}
\end{figure}

Despite the spectral analysis and a large set of astrometric
observations, there were only a few times of minima published
during the last few decades. The main reason is relatively recent
discovery of the photometric variability of HT~Vir. The first
times of minima come from 1979. Since then, there were only 31
observations obtained (see Fig. \ref{HTVirOC}). Four new
observations were obtained. The two of them were carried out in
Ond\v{r}ejov observatory with the 65-cm telescope and Apogee AP-7
CCD camera and 1 second exposure time in R filter. This new times
of heliocentric secondary minima are $2454175.58494 \pm 0.00011$
and $2454195.56170 \pm 0.00007$. The next one was observed by L.
Br\'{a}t (Private Observatory), using 8-cm telescope with ST-8 CCD
camera, 20 seconds exposure time in R filer, resulting in
$2454210.44154 \pm 0.00015$, and the last one by
R.D\v{r}ev\v{e}n\'y with ST-7 CCD camera, 60 seconds in R filter,
resulting in $2454213.49967 \pm 0.00012$. One unpublished
observation by M.Zejda was also used and four times of minima by
M.Zejda published in Zejda (2004) were recalculated, because the
heliocentric correction was wrongly
computed. 

The final plot of the relative astrometric orbit of HT~Vir is in
Fig. \ref{HTVir-orbit}. The results, parameters of the orbit
around the common barycentre of the system, are given in Table
\ref{Table1}. The values of these parameters
($A,\,p_3,\,i,\,e,\,\omega,\,\Omega, T_0, JD_0, P$) were obtained
minimizing the $\chi^2_{comb}$.

Walker \& Chambliss (1985) published the first rough estimation of
the proposed LITE from the astrometric orbit. Their value (0.18 d)
is not too far from ours (0.13 d).

The new elements for the astrometric orbit can be compared to
those of Heintz (1986), which are the following: $p_3~=~274.0$~yr,
$e~=~0.638$, $a~=~1010$~mas, $i~=~42.7^\circ$,
$\Omega~=~176.4^\circ$, $\omega_3~=~250.0^\circ$. As one can see,
the period of the new orbit is a bit shorter, but the main
differences in these values are the angles $\omega$ and $\Omega$.
The same fit to the astrometric data points could be reached with
simultaneously transformed values $\omega_3 \rightarrow \omega_3 +
180^\circ$ and $\Omega \rightarrow \Omega + 180^\circ$. This only
means the interchange of the role of the two components. This
result indicates the incorrect identification of the variable
HT~Vir in the system in our analysis (the variable was supposed to
be the
component A) and also in the WDS catalogue, see WDS notes\footnote{http://ad.usno.navy.mil/wds/wdsnewnotes\_main.txt}. 
While Pribulla \& Rucinski (2006) correctly identified the
variable HT~Vir as the B component and A as a single-lined binary.

If we adopt these parameters to estimate the mass function of the
distant pair (mass function of the whole pair, not the individual
components), we obtain $f(M_3)~=~0.17~$\Mo. This is quite a high
value, dictated by the large amplitude of the LITE. With the total
mass of the primary and secondary $M_{12} = 2.3$~\Mo~we get the
third mass of $M_3 = 2.10~$\Mo. The mass of the distant pair is
quite high (D'Angelo et al. (2006) derived the mass for some 50
per cent lower, $M_3=1.15~$\Mo), but note that also this object is
a binary and we do not know the individual masses. From the
spectroscopic observations (we remind that it is a SB1 type
binary), we are only able to estimate the mass function of the
components, or some upper limit for one of them (we do not know
the inclination). Our resulting $M_3$ is the total mass of the SB1
pair $M_{3,1} + M_{3,2}$; the limit for the invisible-component
mass $M_{3,2}~\cdot~\sin(i) = 0.075~$\Mo. If we assume the
coplanar orbit, high difference in masses would arise, one
component should be much more luminous and also more luminous than
the eclipsing pair itself, which is not the case. In fact the
whole system is not coplanar (see e.g. $i=315.5^\circ$ and the
inclination of the EB close to $90^\circ$). If we assume two
approximately equal masses, there is a problem with the
luminosity, because the distant pair has to be roughly as luminous
as the eclipsing pair. This could only be satisfied if one
component is underluminous or degenerate.

We have to take into consideration also the comment on the
light-curve solution by Walker \& Chambliss (1985). Using the
Wood's model, they discovered that if the third light
$L_\mathrm{3}$ is fixed to be equal to the light from the distant
visual component (it means $L_\mathrm{3}=0.5$), the solution of
the light curve is unrealistic. To conclude, the system could be
much more complicated than the approach we have used here.
Especially because of the resultant mass and luminosity of the
distant pair, the body causing the astrometric variation is
probably different from the one causing LITE, but this conclusion
will be proven only if also the nonlinear part of the $O-C$
diagram is covered.

\section{Discussions and conclusion}

Although the number of systems where the astrometric orbit of the
third component has been measured together with the presence of
LITE is growing steadily, in the most cases only very limited
coverage, both in astrometry and times of minima is available.
Especially due to this reason the combined analysis of these
systems is still difficult, the results are not very convincing
and the resultant parameters are affected by relatively large
errors.

During the last decade a few papers combining the approach of
simultaneous solution of radial velocities, spectral analysis,
astrometry, \emph{Hipparcos} measurements or LITE were published.
Besides the systems mentioned in the introduction (V505~Sgr,
QS~Aql, 44~Boo, QZ~Car, SZ~Cam, GT~Mus, and V2388~Oph) there were
also the analyses of V1061~Cyg (combining the light curve
analysis, radial velocity analysis, light-time effect and
\emph{Hipparcos} measurements, see Torres et al. 2006),
the papers where the radial velocity measurements and astrometry
were combined (see Muterspaugh et al. (2006) for the solution of
LITE system V819~Her, or Gudehus (2001) for $\mu$~Cas), the paper
on HIP 50796 combining the radial velocities with the
\emph{Hipparcos} abscissa data (Torres 2006), or the paper on
$\delta$~Lib comparing the results from the period analysis,
light-curve analysis, spectral analysis, radio emission and
astrometry, respectively; see Budding et al. (2005).

Three eclipsing binaries were studied in this paper. In the case
of VW~Cep, where the orbits in both methods have relatively best
coverage, the distant body satisfies the limits for the
luminosities, and also the systemic velocity variations coincide
with our hypothesis. New results are comparable with the previous
ones. An additional fourth body was introduced to describe the
long-term variation in times of minima. The system is probably
more complicated than we assumed (chromospheric activity cycles,
stellar spots and flares), and we decided to explain only the most
pronounced effects in the $O-C$ diagram. Using this combined
approach it is possible to derive the parallax of VW~Cep more
precisely than in previous papers, resulting in $\pi=(35.85~\pm$
$\pm~0.37)~\mathrm{mas}$. The system $\zeta$~Phe displays an
apsidal motion together with LITE and this explanation fits the
$O-C$ residuals quite well. The quality of the astrometry is
worse, but the main result about the masses and the spectral type
of the third body is in an excellent agreement with the previous
photometric analysis. The last system HT~Vir is the case where the
new resultant parameters of the distant body are about 2 times
larger than one would expect. This could be due to the fact that
only a few times of minima were observed in the linear part of the
$O-C$ diagram, new minima are needed in the next decades.

The final result is that the method itself is potentially very
powerful but it is also very sensitive to the quality of the input
data, especially if the method is used for determining the
distances of these binaries. It can only be applied successfully
in those cases where the astrometric orbit and the LITE in the
$O-C$ diagram are well defined by existing observations.

\section*{Acknowledgments}

This investigation has been supported by the Czech Science
Foundation, grants No. 205/06/0217 and No. 205/06/0304. We wish to
thank Dr. Brian Mason and Dr. Gary Wycoff for sending us the
astrometric WDS data. We wish also thank to Dr. Pavel Mayer and
Dr. Guillermo Torres for helpful and critical suggestions. And
also to Prof. Petr Harmanec for his useful comments and linguistic
corrections. MW wish to thank the staff at SAAO for their warm
hospitality and help with the equipment. This research has made
use of the SIMBAD database, operated at CDS, Strasbourg, France,
and of NASA's Astrophysics Data System Bibliographic Services.


\end{document}